# Modeling Variants of Automotive Systems using Views

Hans Grönniger, Holger Krahn, Claas Pinkernell, and Bernhard Rumpe
Institute for Software Systems Engineering
Technische Universität Braunschweig, Germany

**Abstract:** This paper presents an approach of modeling variability of automotive system architectures using function nets, views and feature diagrams. A function net models an architecture hierarchically and views are used to omit parts of such a model to focus on certain functionalities. In combination with feature diagrams that describe valid variants, the concepts of feature and variant views are introduced to model architectural variants. The relationship between views, variants and the underlying complete architectural model is discussed. Methodological aspects that come along with this approach are considered.

## 1 Introduction

A main challenge in developing automotive systems is the high diversity of possible variants of a system architecture. A modern car consists of a lot of features a customer can choose from which leads to thousands of possible configurations. Additionally, the complexity of automotive systems is increasing rapidly coupled with a growing need for reusability of existing functionality. This is not just a development issue, but it is also getting harder to maintain and evolve running systems. Especially an intuitive notation of system architectures that considers the high number of variants is missing in today's approaches.

This paper focuses on the aspect of variability in automotive system architectures. An architecture description language based on function nets is introduced to model hierarchically structured and logically distributed systems. Hierarchical decomposition allows for handling the complexity of a system architecture and the concept of views is introduced to improve system understanding. In this paper, views are used to describe features of a system. In combination with feature diagrams they also give an overview of possible variations. Variant views are introduced that abstract from a complete function net to the relevant parts of a feature in a certain variant. The paper contributes to the understanding of systems with high diversity by introducing an intuitive notation to model features and their variants.

In Section 2 of this paper the modeling of automotive architectures with function nets is explained as introduced in [GHK+07, GHK+08]. The section also focuses on the definition of views. Section 3 describes what a feature is and how feature diagrams model system variability. The concept of feature views is introduced and variant views are defined. The connection between a feature diagram, feature and variant views is explained by examples. Section 4 considers methodological aspects and in Section 5 related work is presented.

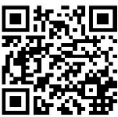

## 2 Modeling Automotive Architectures using function nets and Views

Function nets are a suitable concept for describing automotive logical architectures [vdB04, vdB06]. In our approach, we propose to use a subset of SysML internal block diagrams for function net modeling, as described in [GHK$^+$07, GHK$^+$08]. The next subsection introduces the notation of internal block diagrams. Syntactically, internal block diagrams are valid SysML internal block diagrams [OMG06], but some additional constraints apply as described in [GHK$^+$08]. Internal block diagrams that describe function nets only use directed connectors and no multiplicities. The main advantages of this notation are the intuitive systems engineering terminology and the distinction between the form of a diagram and its use. It is possible to describe both, views and the underlying complete architecture, with the same notation. As a consequence, we achieve a low learning effort and an easy switching of viewpoints between function nets and views. The notation itself should be acceptable for developers from other domains than computer science, because it uses existing systems engineering terminology.

### 2.1 Internal Block Diagrams

Internal block diagrams (ibd) used for function net modeling contain logical functions and their communication relationships [OMG06]. *Blocks* in the diagram can be decomposed into further blocks to structure the architecture hierarchically or can be simple functions. The interface of a block is formed by its input and output channels.

Blocks are communicating over so called *connectors* which are unidirectional and logical signals are sent from a sender to a receiver over them. Connectors can directly connect two blocks ignoring hierarchical composition (cross-hierarchy communication). A signal has exactly one sender, but it may have several receivers.

Figure 1 shows an internal block diagram of a car comfort functionality. The block consists of a *central locking system (CLS)*, the *central locking system control unit (CLSControlUnit)*, doors and a trunk. The *CLSControlUnit* consists of two different push-buttons and a status interpreter.

The CLS receives two external signals: *AutoLockState* and *Speed*. The system outputs *CLSState* to an external environment. A door can be decomposed as seen in Figure 2. It consists of a *DoorContact* and a *LockControl*. As you can see in Figure 1, the function Door is instantiated four times. In our approach the instantiation of blocks is supported which increases reusability of functions and avoids redundant function definitions. More information on the instantiation mechanism can be found in [GHK$^+$07].

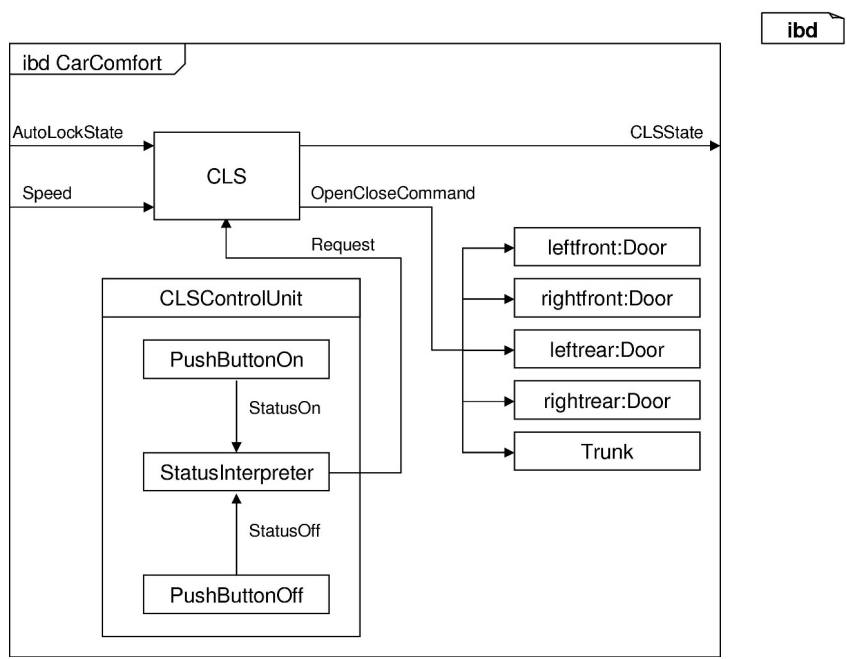

Figure 1: Car comfort function

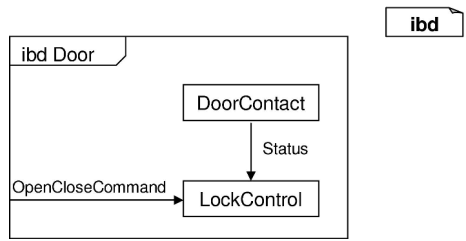

Figure 2: Decomposition of a door

Note that the example is neither complete nor does it illustrate a real design of a comfort function in detail.

## 2.2 Views

Internal block diagrams are also used to define *views* of a given architecture. A view is denoted by the stereotype «view» and helps developers to focus on particular functionality by hiding the complexity of a complete logical architecture. A view has to be consistent to an underlying internal block diagram. That means that a view consists of elements of a complete architectural model.

Views do not introduce new elements but abstract from the whole system by omitting blocks and connectors. An exception is the possibility to model the environment of a block and its non-digital communication. Environmental blocks (depicted by stereotype «env»), like actuators or even "the street" can be modeled. There are several possibilities of non-digital communication, e.g., «M» for mechanical, «E» for analogous electrical or «H» for hydraulic stimulation. First elements for an appropriate theory for at least the analogous and digital communication is given in [SRS99]. Modeling the environment is used for better comprehensibility. It is also possible to import so called external blocks. These blocks are marked with the stereotype «ext» in order to model a view's context in the whole system.

Figure 3 shows a view of the running example. The view depicts a coupe variant of a car comfort function. A coupe has only two doors and a trunk compared to the full model in Figure 1. So the rear doors are omitted.

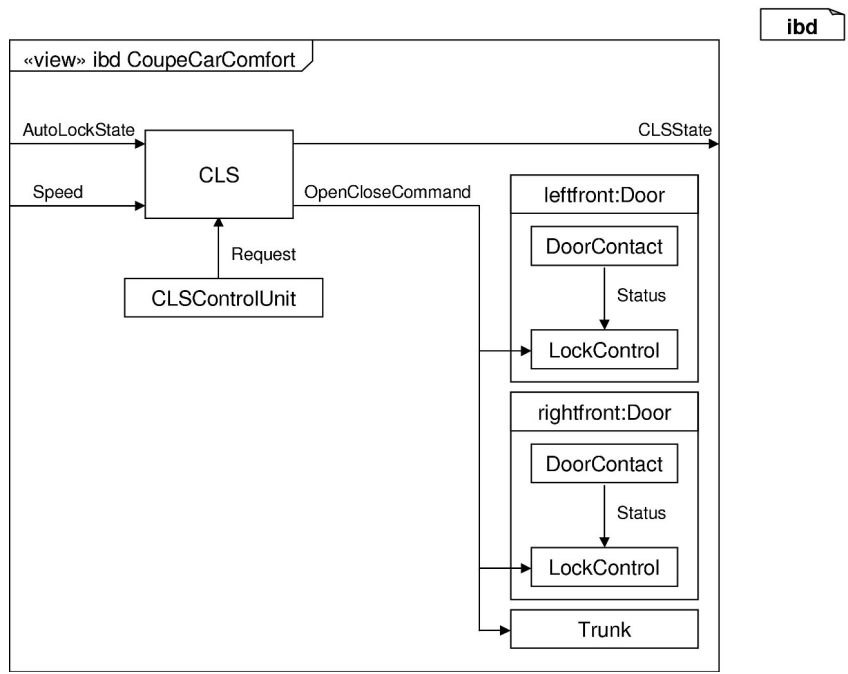

Figure 3: View of car comfort function

Views are abstractions of a complete function net. The following consistency conditions can be checked to validate this relationship [GHK[+]07]:

1. Each block in the view without a stereotype «env» must be part of the complete function net.

2. Whole-part-relationships in the view must be present in the complete function net. However, it is permitted to leave out intermediate layers.

3. Elements that are related via a (possibly transitive) whole-part-relationship in the complete function net must also have this relation in a view if both elements are shown.

4. Communication relationships not marked with a stereotype shown in the view must be present in the logical architecture. If the concrete signal is omitted in the view, an arbitrary signal communication must exist in the complete function net.

5. Communication relationships need not be drawn to the exact source or target. Any superblock is sufficient if the exact source or target block is omitted.

Views on views are also supported by the notation. In that case, an additional context condition applies:

6. A view on a function net is a specialization of another view if both are consistent to the complete function net and the blocks and connectors in the specialization function net are a subset of the blocks and connectors shown in the referred function net.

## 3 Defining Variants with Views and Feature Trees

Our approach of handling variability in automotive systems is based on the idea to use views to describe features and their variants [CE00]. We first recall feature diagrams as a standard way to model the feature variability and possible valid configurations of a system. Next, we define feature views that are used to describe the logical architecture of a particular feature. Subfeatures of a feature are defined as views on the parent feature's view. Variant views can be calculated from the views defined for subfeatures and their relationship expressed in the feature diagram. The connection between the complete logical architecture, a feature diagram, feature views and induced variant views is depicted in Figure 4.

A view of the complete function net is defined for feature F (dashed rectangle). Subfeature views for S1 and S2 are the (overlapping) views indicated by the dotted rectangles. The or-relationship in the feature diagram induces an additional variant "S1 and S2" marked by the solid rectangle.

### 3.1 Features

A *feature* is an observable unit of functionality of a system. Especially in automotive systems a customer can choose from a variety of features. So there is a multitude of possible feature combinations that make up the complete automotive system. A feature might be a single function or a subset of interacting functions of a complete function net. These functions can also be shared between distinct features.

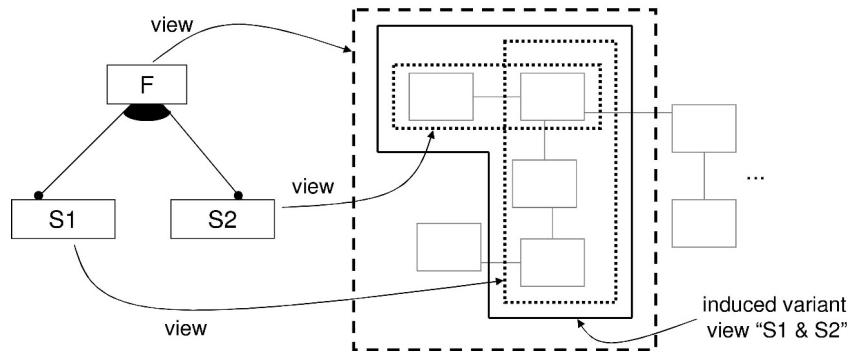

Figure 4: Relationship between the complete logical architecture, its feature views, and their variant views.

Features with subfeatures are usually organized in so called *feature diagrams* (fd) as explained in [CE00]. In our approach we use this notation in addition to function nets and feature views that are described below. A feature diagram, as we use it, is a tree structure with features as its nodes. A feature might be optional (depicted by an empty circle) or mandatory (depicted by a black circle). Features can have a set of alternative subfeatures, depicted by an arc that subsumes all possible alternatives. That means that exactly one subfeature from the set of alternatives is used. A filled arc is used to describe or-features meaning that every combination of subfeatures is allowed. Edges without an arc are used to describe that a feature consists of several subfeatures.

Figure 5 shows an example feature diagram. The car consists of a mandatory feature Engine and an optional feature Comfort Functions. The engine can be chosen from Gasoline, Electric or Hybrid. Any combination of Navigation System, Air Condition and Central Locking System yields a valid Comfort Functions feature.

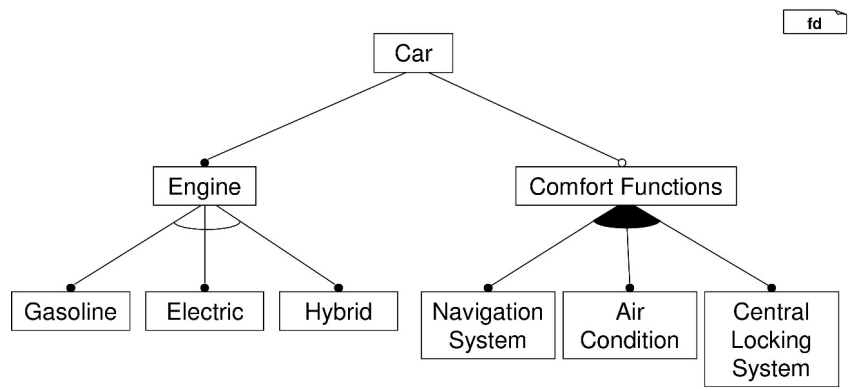

Figure 5: Feature diagram for a simple car

## 3.2 Feature and Variant Views

As indicated in Figure 4, the (structure of a) feature is defined as a view on a complete function net. This has the advantage that a single feature can on the one hand be understood and analyzed in isolation. On the other hand its embedding in the relatively complex complete function net is explicitly modeled. Views describing features are called *feature views*.

In views, parts of a complete function net that are not relevant for a certain feature are omitted. A feature view only shows relevant blocks and signals. But it may also contain physical devices and even the environment of the car. Figure 6 shows a door with its environment. As explained in Section 2.2, a feature view has to be consistent to a complete function net which can be checked automatically.

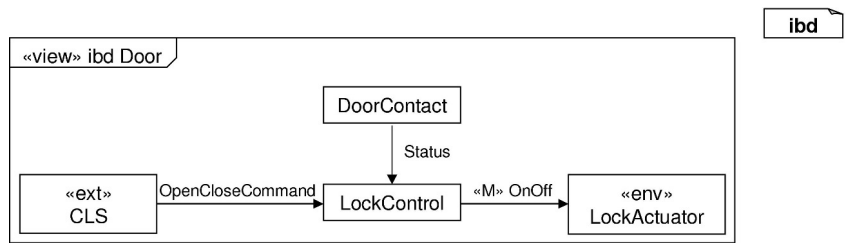

Figure 6: View of a door with its environment consistent to Figure 2

The underlying complete function net that describes the whole logical system architecture consists of all features in all possible variants. The so called complete model or "150 percent model" is an important concept of our approach. The logical architectures of automotive systems are usually 150 percent models. But the realization of the logical architecture as ECUs is normally not a 150 percent model. Still, the realization often contains unused functionality, because it seems more economic to design every car with the same set of basic functionality even if it is not used in the concrete configuration. From a 150 percent model variants can be chosen by parametrization. A disadvantage of this approach is a complex logical architecture. The complexity is handled by views. Each (sub)feature is described by one *view*, c.f. Figure 4. Please note that subfeatures are defined as views on feature views. In that case, the last context condition from Section 2.2 has to hold, i.e., views may not introduce new functions or communication relationships compared to the underlying feature view.

The set of possible variants and according views is calculated from the (sub)feature views and their relationship in the feature diagram which is explained in detail in the next section.

## 3.3 Variability with Feature Diagrams and Views

We model several feature diagrams and consistent feature and variant views. Note that in the following examples the complete function nets are left out. The given feature views

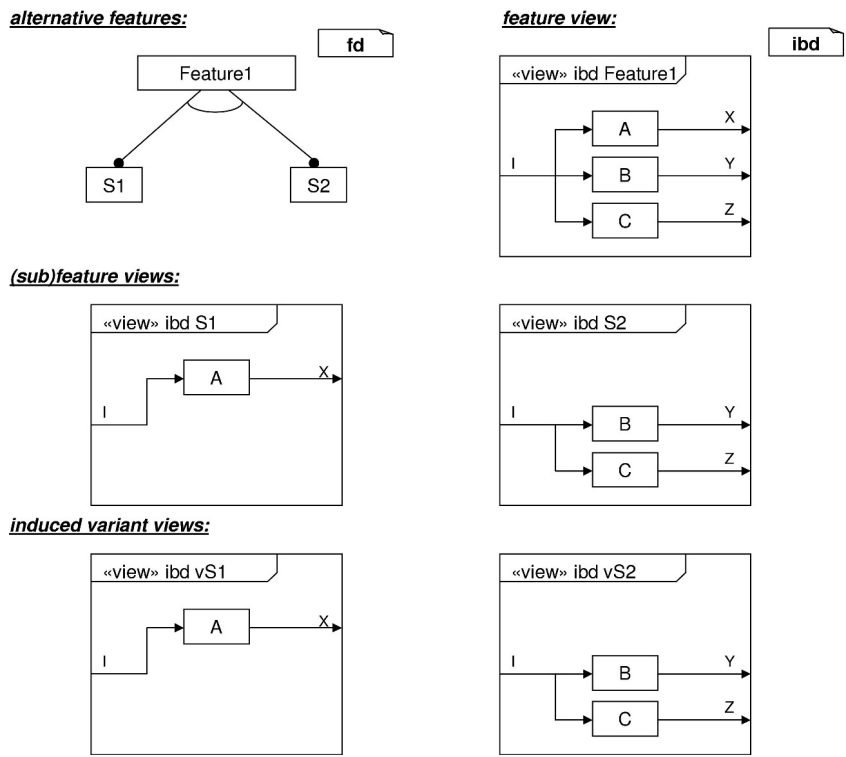

Figure 7: Alternative features

contain the relevant information for the examples.

**Alternative Features**

Figure 7 shows a feature with its two alternatives subfeatures S1 and S2. The figure also depicts the underlying (150 percent) feature model that is the basis for the two alternatives. Also, all possible variants are shown which, in case of alternatives, coincide with the subfeature views.

We assume that each block has an incoming signal I. Block A has an outgoing signal X, and so on. As motivated later, the complete model still has to be a valid function net in which each signal has exactly one sender, so sharing outgoing connectors is not allowed. Some outgoing signals are only used in one variant. Now, this could mean two things, a) the signals actually have the same signal type and semantics but since it is not possible to have the same signal name with multiple senders, different signal names have to be invented, or b) the signals are not connected in any way. While b) is not a problem, in case a) one could argue that the logical architecture should preserve this information. This could, for instance, be achieved by adding another logical block to the architecture that receives three distinct signals and delegates one common signal. Another possibility

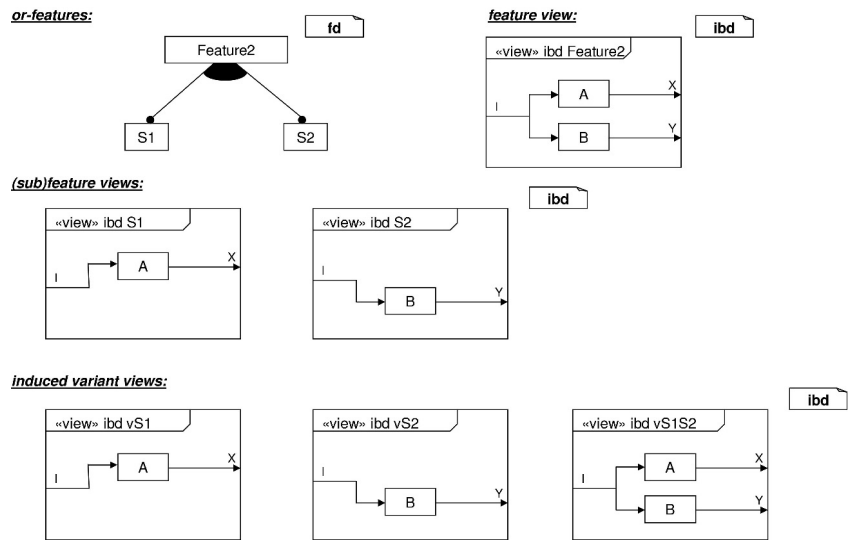

Figure 8: Or-features

would be to add signal types and instances to the function nets.

**Or-Features**

Figure 8 depicts a feature that may consist of subfeatures S1 or S2 or both of them. As in our previous example we assume that we have an incoming signal I and outgoing signals X, Y. In variant vS1S2 both subfeatures (S1 and S2) are active. So both, X and Y, are sent by Feature2. Again, if X and Y are only different names for (conceptually) the same signal, the receiving blocks or functions have to consider this. In that case, if both signals target at the same function, additional mechanisms (e.g., an arbiter) could be inserted.

**Mandatory and optional features**

In addition to alternative and or-features we have the composition of subfeatures. In this situation, a subfeature can be mandatory or optional. Figure 9 shows an optional feature S1 and a mandatory feature S2. We assume that we have I as an incoming signal. Block A has an outgoing signal X and B has an outgoing signal Y. The given feature diagram implies two valid variants as seen in our variant views: The variant in which only B is active and another variant in which A and B are active.

Note that there are more combinations of features possible. The basic elements of the notation were introduced here. In [CE00] the normalization of feature combinations is explained in detail.

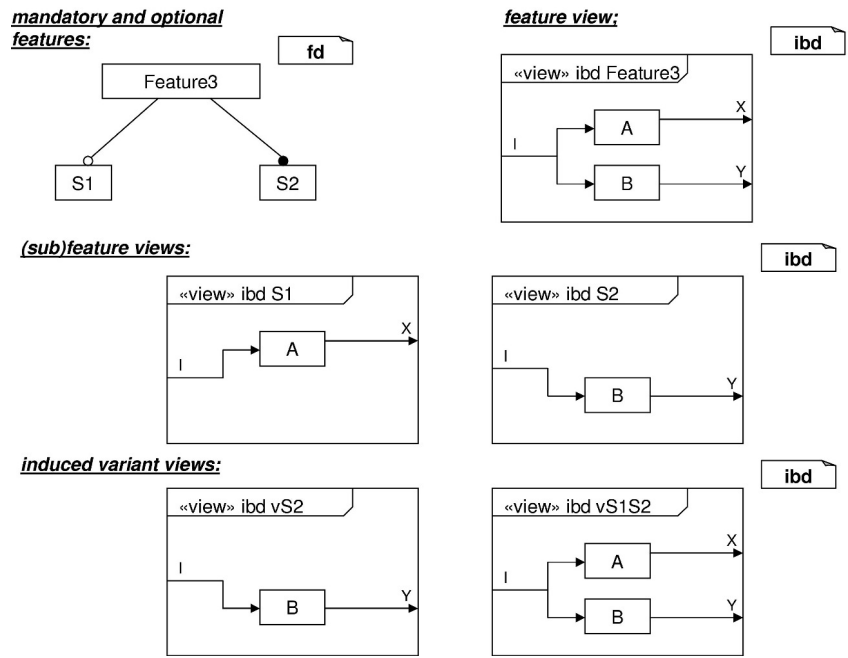

Figure 9: Mandatory and optional features

## 4 Methodological Aspects

View-based development of an automotive logical architecture can be integrated in a typical automotive development process as described in [GHK+07, GHK+08]. Some additional methodological aspects are discussed below.

**Developing Automotive Systems**

With this approach we want to introduce an extended methodology and tool chain, that uses an intermediate layer to describe the logical architecture using function nets. This intermediate layer in between the requirements and the implementation phases improves the understandability of automotive systems by focusing on feature modeling and increases reusability.

The features of a system are extracted from requirements that were captured using a requirements management system, e.g., DOORS [Tel08]. The logical architecture is developed by using function nets. The features from requirements are mapped to the logical architecture by feature views. For a better comprehensibility and to abstract from the complete "150 percent model", feature diagrams and variant views are used. In subsequent development steps the logical architecture can be mapped to a realization, e.g., using the

AUTOSAR [Aut08] methodology.

Automotive systems are not developed from scratch. A new generation of a system is usually based on its predecessor. Being able to reuse large parts of a logical architecture (not just implementations as proposed by AUTOSAR) is hence a key requirement. This is supported by our approach in that the complete logical architecture and feature views can be used as a starting point for the development of the next vehicle generation. Changes are made to the views to fullfil new requirements. Consistency to the complete model is then checked and possible conflicts and inconsistencies can be tracked down automatically and resolved manually since this in general is a complex engineering task.

We hence propose the following development steps to model a logical architecture:

1. Create or adapt feature diagrams.

2. Reuse the complete logical architecture and views from an earlier vehicle generation.

3. Develop self-contained views to describe new features.

4. Adapt feature views to reflect changes in the requirements specifications.

5. Early and repeatedly execute consistency checks between views and the complete function net of a predecessor to identify inconsistencies or unused blocks in order to modify the complete function net.

**Advantages and Limitations of a 150 percent model**

When developing complete systems, like cars, we can safely assume that at some point the complete system architecture is developed including all possible variations. The so called 150 percent model of an architecture constitutes the basis for modeling system configurations with views. Especially in the automotive domain it is not feasible to generate and test every possible variant configuration. As a consequence, analysis, verification, and testing of an automotive logical architecture can much better be carried out on this 150 percent model that integrates all variants. Therefore, we need the complete logical architecture to be a valid function net (e.g., with one sender per signal only). This however does not mean that we cannot run tests on incomplete models or views. Quite the contrary: It is mandatory to clarify certain functional and non-functional properties early on views and reuse these tests for complete function nets.

**Top-down vs. bottom-up development**

The development process in general could be a top-down, a bottom-up process or a combination of both. In a top-down process the logical architecture is modeled on a high level of abstraction and later each element is refined until it has the desired structure. In a bottom-up process modeling stays very close to an implementation. The concept of views can be used to hide the complexity of such an architecture. Today, in the automotive sector most software is designed from bottom-up. One reason is that the placement of a logical function is often already decided because the ECU that is realizing this function is already predetermined. Reuse of system components is not limited to the logical architecture, but is also an aspect on the hardware level. To consider already fixed design decisions, function nets can be equipped with attributes that document these decisions early in the development process [GHK$^+$07].

However, a stringent use of views on function nets can lead to a substantial change of method, as it will be possible to model functional units and assemble them at will. Thus function nets can be developed top-down, bottom-up or even starting in the middle. Furthermore, it will be possible to evolve function nets using calculi like [KR06, PR97, PR99]. Consequently, modeling architectures with function nets and views supports both, top-down and bottom-up modeling. In practice, we assume a mixture of both where new requirements are propagated top-down through the logical architecture that is reused "bottom-up" from an earlier vehicle generation.

## 5 Related Work

In [CE00] the modeling of features is described using a tree-based notation. This notation is widespread and intuitive. This paper adapts the notation of feature diagrams to model features of automotive systems.

An important concept of handling variability in software systems is introduced in [PBvdL05]. The authors explain how variability of system can be handled as so called product lines. A notation to model the variation points and the variants of a system is introduced. We decided to follow the notation in [CE00] because in our opinion it is more intuitive and often used.

An approach to model the logical architecture of automotive systems with UML-RT is introduced in [vdB04]. This approach is extended by using internal block diagrams from SysML [OMG06] to model logical architectures and views.

In [SE06] the possibility of merging views to a complete system are discussed. Merging views to a complete system could be interesting in a strict top-down development process. Nevertheless, we do not propose to merge views because of constant evolution of requirements. Checking consistency between views and a complete function net eases finding inconsistencies that have to be resolved manually.

In AML [vdBBFR03, vdBBRS02] functions are tailored by introducing a variant dependency in class diagrams. AML uses a strict top-down design of functions. In our approach, logical functions can be shared across multiple features and also variants.

The automated derivation of the product variants of a software product line are explained in [BD07]. The derivation is illustrated by an example product line of meteorological data systems. As notation a simplified variant of feature diagrams from [CE00] is used.

AUTOSAR [Aut08] is a consortium that standardizes the architecture of automotive systems and improves the reuse of software components. Our approach proposes the AUTOSAR methodology to realize a given function net. But we think it is necessary to have an intermediate layer in between requirements and AUTOSAR to improve the understandability and the reuse of automotive features. Our approach of modeling logical architectures with function nets fills out this gap.

## 6 Conclusion

In [GHK$^+$07, GHK$^+$08] a notation to model logical architectures as function nets is introduced that fills the gap between requirements and the realization of automotive systems. Based on that notation, the concept of views was explained using the same notation. In this paper the concept was extended to model variants of an architecture. In combination with feature diagrams, feature views and variant views are used to communicate variability of automotive system architectures.

Especially in developing software for automotive systems the explicit modeling and an efficient handling of the high diversity of configurations is important. The variability of a system is mapped from requirements to the logical architecture using views. From there the variability can be mapped on a concrete realization, for instance with help of the AUTOSAR methodology.

Generally, tool support is a fundamental aspect of modeling architectures. So, in future work the tool support has to be extended. With the extension of the tool chain, we think our approach helps to handle variability and complexity of automotive systems in existing development processes. In the future, we also will investigate how the approach can be extended by models that describe the behavior of a feature.